**Title**

Exploring the dynamic interplay of cognitive load and emotional arousal by using multimodal measurements: Correlation of pupil diameter and emotional arousal in emotionally engaging tasks

**Authors**

*Kosel, C. Friedl-Schöller Endowed Chair for Educational Psychology, TUM School of Social Sciences and Technologies, Technical University of Munich (TUM), Munich, Germany

Mail: christian.kosel@tum.de
ORCID-ID: 0000-0002-4857-4148

*Michel, S. Assistant Professorship for Business, Economics and Vocational Education, TUM School of Social Sciences and Technologies, Technical University of Munich (TUM), Munich, Germany

Mail: selina.michel@tum.de
ORCID-ID: 0009-0003-6014-708X

Seidel, T. Friedl-Schöller Endowed Chair for Educational Psychology, TUM School of Social Sciences and Technologies, Technical University of Munich (TUM), Munich, Germany

Mail: tina.seidel@tum.de
ORCID-ID: 0000-0002-2578-1208

Förster, M. Assistant Professorship for Business, Economics and Vocational Education, TUM School of Social Sciences and Technologies, Technical University of Munich (TUM), Munich, Germany

Mail: manuel.foerster@tum.de
ORCID-ID: 0000-0001-8125-4595

*These authors contributed equally to the manuscript

**Abstract**

Multimodal data analysis and validation based on streams from state-of-the-art sensor technology such as eye-tracking or emotion recognition using the Facial Action Coding System (FACTs) with deep learning allows educational researchers to study multifaceted learning and problem-solving processes and to improve educational experiences. This study aims to investigate the correlation between two continuous sensor streams, pupil diameter as an indicator of cognitive workload and FACTs with deep learning as an indicator of emotional arousal (RQ 1a), specifically for epochs of high, medium, and low arousal (RQ 1b). Furthermore, the time lag between emotional arousal and pupil diameter data will be analyzed (RQ 2). 28 participants worked on three cognitively demanding and emotionally engaging everyday moral dilemmas while eye-tracking and emotion recognition data were collected. The data were preprocessed in Phyton (synchronization, blink control, downsampling) and analyzed using correlation analysis and Granger causality tests. The results show negative and statistically significant correlations between the data streams for emotional arousal and pupil diameter. However, the correlation is negative and significant only for epochs of high arousal, while positive but non-significant relationships were found for epochs of medium or low arousal. The average time lag for the relationship between arousal and pupil diameter was 2.8 ms. In contrast to previous findings without a multimodal approach suggesting a positive correlation between the constructs, the results contribute to the state of research by highlighting the importance of multimodal data validation and research on convergent vagility. Future research should consider emotional regulation strategies and emotional valence.

## 1. Introduction

The assessment of learners' cognitive and emotional states has long been a focus of educational research, with the primary goal of optimizing the learning experience and fostering performance (Moon et al., 2020). Cognitive and emotional states play a central role in determining learning effectiveness, as they significantly influence a learner's attention span, information processing, and holistic performance (Kärner et al., 2016; MacDougall et al., 2013). Considering learning and problem-solving as multifaceted processes involving interaction between cognitive and emotional states (Noorozi et al., 2019), multimodal measurements using state-of-the-art sensor technology to assess cognitive and affective processes in real-time are promising to analyze learning in a more holistic way and improve individualized and adaptive learning activities (Mayer et al., 2023; Cloude et al., 2020).

Traditional methods, particularly self-report measures, have frequently been employed as a prevalent tool for evaluating cognitive and emotional states (Schmidt-Weigand & Scheiter, 2011). For example, learners may be asked to rate the perceived difficulty of a task or to describe their emotions after completing a learning activity (Schmidt-Weigand et al., 2010). Sensor technology such as eye-tracking, emotion recognition, heart rate detection or pupillometry now allows researchers to collect real-time data from different data streams at the same time (Cloude et al., 2020). This opens up opportunities for multimodal measurements to analyze more than one indicator and allow to study interactions between cognitive and affective states and regulation processes in multifaceted learning processes. Importantly, there is an observable shift from reliance on self-reported measures to the incorporation of objective or direct measurement techniques. This transition highlights the differentiation between subjective perceptions and externally observable data, enabling a more precise and unbiased analysis of learning behaviors and outcomes. Such objective measures can provide insights into the actual dynamics of learning processes, beyond what learners may report about their own experiences. This methodological evolution underscores the importance of leveraging diverse measurement approaches to capture the complex interplay between cognitive functions and emotional states in education (Noorozi et al., 2019).

These multimodal approaches are emerging in educational sciences (Noorozi et al., 2019). For example, Dubovi (2022) demonstrated that 51% of pre-post knowledge achievements can be explained by a combination of modalities measured by different psycho-physiological sensor streams. However, Noorozi et al. (2020) identified that out of 207 multimodal publications only 14 publications used cognitive and emotional states at the same time and only ten publications used a combination of cognitive, motivational and emotional states. One reason might be that there are several challenges in multimodal research like challenges in data integration and analysis across data channels (e.g., different sampling rates), the uncertainty about the amount of data that needs to be collected to investigate the underlying processes or the impact of transformation from raw data to interpretable and actionable data, which is often done by data aggregation or filters (Azevedo et al., 2018).

In the field of multimodal learning analytics research, several questions concerning data integration emerge (Mu et al., 2022). According to Mu et al. (2022), it is crucial to ask which type of data is suitable for learning indicators, how multimodal data can be aligned ensuring that all indicators are well reflected and how to consider complementary correlations from inter- and cross-modality effectively. These correlations are sometimes hidden and neglected (Mu et al., 2022). Therefore, "many to one" approaches for multimodal data integration to improve the accuracy of measurements and "many to many" approaches to multimodal data integration for improving information richness might feed in "mutual verification between multimodal data" for empirical evidence on data fusion and integration (please see Mu et al., 2022 for further review). Multimodal data validation and verification also allows to analyze relationships and correlations between different data streams and indicators (Mu et al., 2022). Taking a closer look on these relationships and correlations also raises the question of concurrent and congruent validity of data streams, which is considerably important for educational practice. Psycho-physiological sensor streams such as heart rate and heart rate variability (Delliaux et al., 2019), electroencephalogram (brain wave levels) (Yoo et al., 2023), electrocardiogram (Chanel et al., 2019), galvanic skin response (Elahi & Islam, 2019), and respiration (Grassmann et al, 2016) share one notable limitation: the need for participants to be physically present in a laboratory setting and to wear specialized hardware, suggesting that these psycho-physiological sensor streams are not easily transferable outside of the laboratory. Recent technological advancements have catalyzed a paradigmatic shift towards digital and remote technology-enhanced learning, thereby significantly elevating the relevance and appeal of psycho-physiological sensor streams beyond traditional laboratory settings (Chernikova et al., 2020).

One way of assessing learners' cognitive load in remote digital learning environments is to analyze their log file data streams, for example, how much time they spent on tasks, the number of attempts on a problem, or patterns of resource access (Hulshof, 2005). However, emotions are hard to assess using log-file methods (Janning et al., 2016), researchers recognized this challenge and have initiated endeavors to incorporate psycho-physiological measures, extending their applicability to remote contexts beyond the confines of laboratory settings (Chanel et al., 2019). Therefore, psycho-physiological measures which can be used remotely, like pupillometry or emotion recognition move into focus. To sum up, it is key to reflect on multimodal data integration by considering intercorrelations between indicators for cognitive and affective states to enhance data economy and practical usability without losing the benefits of multimodal measurements described above. In this work, we focus on investigating systematic relationships between two established indicators highly relevant for the educational context: Pupil diameter as an indicator for *cognitive load* and a emotion recognition system based of the Facial Action Coding System (FACS) with deep learning for *emotional arousal*.

1.2 Cognitive Load

"The level of an individual's measured effort in order to cope with one or more cognitively demanding tasks" can be applied as a definition for cognitive load (Skarmagkas et al., 2021, p.2). Cognitive load is closely related to constructs such as mental workload, mental effort, or mental demand in ergonomics or human factors literature which also describe spending cognitive resources and mental activity required to perform a task (Vanneste et al., 2020; van de Acker et al., 2018; Young et al., 2014). To make a more precise differentiation, cognitive load incorporates demands from the learning environment to perform learning tasks [Task-driven mental load], as well as the more human-centered dimension of actively spending cognitive resources while processing the learning task in terms of cognitive engagement [mental effort] (Schnaubert & Schneider, 2022; Pass & van Merriënboer, 1994). The constructs of mental load and cognitive load are often used interchangeably (Schnaubert & Schneider, 2022), whereas the construct of cognitive load is more common for educational sciences. Since we do not aim to differentiate between task-driven mental load and mental effort, we decided to focus on the concept and term of cognitive load within this study. Cognitive-load researchers traditionally aim to assess learners' cognitive functioning for instructional adaptations because optimal learning environments or tasks should neither cause cognitive over- nor underload (Martin et al., 2021; Sweller et al., 1998). Research centered on cognitive load theory emphasizes the finite capacity of working memory, as opposed to the expansive storage potential of long-term memory (Paas, 2014; Sweller et al., 1998). Evidence suggests that flooding working memory with too much information can lead to cognitive overload, impairing the brain's ability to efficiently process and assimilate the information presented (Sörqvist et al., 2016). Commencing with the seminal work of Löwenstein in 1920, one stream of research—pupillometry—is concerned with small changes in pupil diameter that are attributed to reflect changes in human cognition (Van Der Wel & Van Steenbergen, 2018; Löwenstein, 1920). Following from this, when assessing cognitive load, pupil dilation is a well-researched indicator. Studies on task-evoked pupillary responses focus on how changes in cognitive load during a task can be measured through changes to pupil size compared to baseline. According to Mallick et al. (2016) pupil dilation showed a positive relationship with cognitive workload in dynamic and unconstrained tasks. Soussou et al. (2012) also reported coherence for the measure of pupil dilation and EEG-based gauges as indicators for cognitive workload. Rodemer and colleagues (2023) used pupil dilation as a cognitive load measure in instructional videos on complex chemical representations and found correlations between pupil dilation and extraneous cognitive load. Concurrently, the advent of flexible open-source platforms dedicated to high-resolution pupillometry, stemming from vision research (i.e., Zandi et al., 2021), presents a promising direction for researchers exploring the incorporation of pupillometry within remote technology-enhanced learning modalities in order to assess cognitive load. While there has been extensive research on cognitive load in isolation (Antonenko et al., 2010), recent findings suggest that a learner's emotional state, especially the learner's level of emotional arousal can influence learning by altering cognitive load (LeDoux, 2021; Tyng et al., 2017).

## 1.3 Emotional Arousal

Emotional arousal, defined as "a state that describes the level of calmness (i.e., low arousal) or excitation (i.e., high arousal) elicited by a stimulus" (Skaramagkas et al., 2020, p. 2), is one bipolar dimension used to group emotions in multidimensional scaling (Russel, 1980) and, thus, proves to be a higher-level dimension for describing emotional states. The rapidly growing methodology in facial recognition offers another promising avenue, providing nuanced real-time analysis for emotion detection (Lewinski et al., 2014). One of these emerging methods is called Facial Action Coding System (FACS) with Deep Learning. The FACS categorizes and describes all the conceivable facial muscle movements (termed "action units" or AUs) and their visible effects. FACS is often used in psychology to objectively describe facial expressions, which can then be associated with specific emotions or states (Lewinski et al., 2014). It allows high-frequent and unobtrusive observations of changes in student emotion during learning and problem-solving situations (Dubovi, 2022; Dindar et al., 2020). Combining FACS with deep learning techniques, such as convolutional neural networks (CNNs), can provide a powerful approach to extracting and interpreting complex facial muscle actions to automatically recognize and classify facial action units from images or video feeds (Bhattacharyya et al., 2021). This automation can allow for real-time emotion detection, which has applications in various fields like human-computer interaction (Podder et al., 2023) and psychology (Gao & Ma, 2020; Song, 2021). Current research suggests that FACS with Deep Learning is comparable to electromyography (EMG) data (Kulke et al., 2020) and self-report data (Harley et al., 2014). To summarize, in multimodal measurements emotional arousal can be captured using FACS with deep learning.

## 1.4 Interplay of Cognitive Load and Emotional Arousal

A variety of findings suggest a dynamic interplay between cognitive load and emotional arousal. On the one hand, emotional arousal not only acts as a source of extraneous cognitive load (Mehrotra & Gunalakshmi, 2020) but also plays a role in affecting the human memory system (LeDoux, 2021). When learners are in a state of high emotional arousal, some of their cognitive resources are directed towards processing these emotions, leaving fewer resources available for other tasks and, hence, increasing the cognitive load. According to the Yerkes-Dodson Law, there is an optimal level of emotional arousal that supports learning; levels that are too low or too intense can hinder the learning process (Sherwood, 1965; Yerkes & Dodson, 1908). On the other hand, cognitive load can also affect emotional arousal (Jerčić et al., 2018). A task that imposes high cognitive load can induce emotions such as frustration or anxiety, leading to an increase in emotional arousal. For example, if a learner finds a task too difficult or complex, this could induce feelings of stress or anxiety, which would result in heightened emotional arousal (Jerčić et al., 2018). Current research supports this outlined reciprocal relationship between cognitive load and emotional arousal, with changes in one often leading to changes in the other (Tyng et al., 2017). Neurobiological research indicates that changes in pupil diameter can be attributed to the activation of the locus coeruleus within the noradrenergic system. This activation, often precipitated by emotional arousal, is postulated to have a consequential role in the process of memory consolidation (Van Der Wel & Van Steenbergen, 2018). Subsequently, several studies have confirmed that pupil dilation serves as a credible biomarker of cognitive load and emotional arousal, highlighting its diagnostic potential in various research contexts (Bradley et al., 2008; Steinhauer, 1983). Although a growing number of studies have used pupil diameter to indicate the level of arousal (Ebitz & Platt, 2015; Kreuzmair et al., 2016; Murphy et al., 2014; Nassar et al., 2012), research that directly examines the relationship between pupil diameter and emotional arousal both measured with is psycho-physiological sensor streams is limited. For example, Kreuzmair and colleagues (2017) found that participants' pupil diameter was a measure of their emotional arousal in hypothetical medical scenarios. In their experiment, they varied the "risk level" in their scenarios as a proxy for different levels of arousal, however, they aggregated data to mean values for specific times of interest instead of using process data based on real-time psycho-physiological sensor streams of emotional arousal. This also reflects the challenges of multimodal research described above. Therefore, currently, there is very limited evidence about how pupil diameter and emotional arousal correlate in time series data. To address this research desiderate, different methodological considerations need to be undertaken.

Considering this interplay between cognitive load and emotional arousal, both high-resolution pupillometry and face recognition based on deep learning techniques have a high potential to be significant methods for assessing learners' cognitive load and emotional arousal within technology-enhanced remote learning modalities (Cloude et al., 2022; Azevedo et al., 2018). However, it is noteworthy to highlight that both data sampling methods use different indicators that do not intersect, and questions of data synchronization and integration arise. While high-resolution pupillometry focuses on the variation of the pupil diameter, face recognition uses complex facial muscle actions to classify emotional states, however, ignoring the eyes (Lewinski et al., 2014). Both psycho-physiological data streams aim to capture facets of cognitive load and emotional arousal using different data streams. Consequently, within an individual, these streams should exhibit a substantial correlation. From a perspective of multimodal data validation (Mu et al., 2020), the validity of these streams as reliable biomarkers must be questioned if there is no correlation between the two data streams. Therefore, the correlation between the two streams of data remains a significant question. Furthermore, the time lag between emotional arousal and physiological responses, such as pupil diameter, is largely unclear. Generally, emotional responses are described to precede physiological responses (Mauss & Robinson, 2009). However, there is currently no concrete evidence detailing the time lag between these two physiological indicators, especially in relation to their respective sampling methods. For multimodal data validation and practical use in, for example, adaptive learning systems, we need to understand if both multimodal data streams for cognitive load and emotional arousal correlate and which stream first shows an indication that might be contextually relevant for feedback or interventions in the future. This is also important considering the analytic bottleneck described by Azevedo & Gasevic (2019) that occurs when merging data streams in educational practice and can lead to latencies in delivering cues and inferences, which might in turn also have negative effects on learning. Therefore, it is key to also reflect on time lags between correlating indicators and consider first indications as "initial reactions".

1.5 Everday Moral Dilemmas as Emotionally Engaging and Cognitively Demanding Tasks

The dynamic interplay between cognitive load and emotional arousal may occur in different contexts where learning and problem-solving take place. One problem-solving context, which is cognitively demanding and emotionally engaging, are everyday moral dilemmas. Dilemmas are short stories or vignettes presenting a moral conflict that confronts a person with two incompatible courses of action leading to consequences that represent rival moral principles or values (Christensen & Gomila, 2012). These moral conflicts can be conflicts between personal interests and accepted moral values, different duties, a set of apparently incommensurable values, or conflicts stemming from one unique moral principle (Christensen & Gomila, 2012). We argue that everyday dilemmas, by their very nature, require individuals to weigh multiple factors, outcomes, and personal values. Typical sacrificial moral dilemmas can be criticized for showing a lack of external validity and realism (e.g. Baumann et al., 2014; Kahane, 2015), this study uses three everyday dilemma situations from a work context as the emotional dilemma task. Everyday dilemma situations can also be cognitively demanding and stimulate emotions (e.g. Thomson & Berenbaum, 2006). For example, goal or role conflicts in everyday work situations such as moral conflicts are described to be challenging for employees and can trigger emotions (e.g. Fisher & Ashkanasy, 2000). Research on moral judgement is continuously discussing the role of emotions in the moral evaluation process (e.g. Blum, 2023; Greene at al., 2001, 2009). One focus is to investigate the role of strong emotional reactions, working as a kind of alarm bell, stimulated by personal dilemma as described in the dual process theory (see Greene et al., 2001, 2009 for further review). According to this, an analysis of the interplay between cognition and emotions is also especially interesting for intervals where high arousal is perceived. To sum up, the complexity of everyday dilemmas can lead to an increase in cognitive load as one's brain processes the information and contemplates potential consequences which also affects emotional arousal (Cummings & Cummings, 2012; Greene et al., 2009; Bluhm, 2013). Therefore, we argue that everyday moral dilemmas are suitable to be used as an emotionally engaging problem-solving task. Using text-based dilemma might also held the advantage that previous research already showed that they are suitable for studies using psycho-physiological sensors like eye-tracking (Garon et al., 2018; Fiedler et al., 2013; Gaffari & Fiedler, 2018; Pärnaments et al., 2015) and emotion recognition (Cumming & Cumming, 2012; Valdesolo & DeSteno, 2006; Gleichgerrcht & Young, 2013).

In summary, based on the evidence described above, we argue that pupil size as an indicator of cognitive load (e.g., Van Der Wel & Van Steenbergen, 2018; Mallick et al., 2016; Soussou et al., 2021) and emotional arousal analyzed by FACS with deep learning (e.g., Bhattacharyya et al., 2021) correlate, particularly in an emotionally engaging task. This correlation as part of the multimodal validation (Mu et al., 2020) can provide starting points to improve learning environments in, for example, adaptive learning settings using only one of the two indicators to assess cognitive and emotional states as a cue for adaptations (Sailer et al., 2023; Chanel et al., 2019). Thereby, also questions of concurrent validity can be addressed. To understand whether this is a reliable perspective for learning interventions, a process-oriented approach analyzing the correlation between two continuous psycho-physiological data streams, pupil size operationalizing cognitive load and FACTS analysis to operationalize emotional arousal, is needed.

## 2. Aim of the Study

The aim of this research is to synchronize pupil diameter and emotional arousal during emotionally engaging problem-solving tasks such as everyday moral dilemmas and to examine whether and how these two measures (i.e., the psycho-physiological data streams) correlate. Thereby we aim to propose a methodological approach for describing the interplay of cognitive and emotional processes. We postulate that synchronizing these data streams would allow us to delineate the dynamic correlation between cognitive and emotional processes, thereby going beyond studies that use either pupillometry or emotion recognition and no multimodal measurement in the context of moral dilemma tasks (e.g., Doerflinger & Gollwitzer, 2020; Fiedler et al., 2013; Garon et al., 2018). Given the research gap, the current study will focus on the research question described below:

**RQ1a:** Can changes in emotional arousal, as captured through facial expressions, be observed concurrently with corresponding variations in cognitive load, as indicated by changes in pupil diameter?

First, we aim to explore the relationship between emotional arousal and cognitive load, as proposed in our study by synchronizing pupil diameter, collected by screenbased eye-tracking technology (Tobii Spectrum Pro), and emotional arousal data from the Noldus FaceReader, using FACS and deep learning (Lewinski et al., 2014). It identifies and classifies facial muscle movements to determine an individual's emotional state. We handle both data streams as classical synchronized time-series (Liu, 2004). By analyzing the correlation of these two streams, we aim to understand whether changes in one modality coincide with corresponding changes in the other. This exploration will shed light on the potential interconnectedness of emotional responses captured through facial expressions and physiological measures and, thereby, contributing to multimodal data validation of and questions about convergent validity of the two data streams. Beyond that, correlation for specific epochs based on research in learning and moral problem-solving is of interest.

**RQ1b:** How does pupil diameter correlate with emotional arousal during epochs of high, medium, and low emotional arousal?

Secondly, contributing to research on the role of strong emotional reactions in personal moral dilemma situations (e.g. Greene et al., 2001, 2009), examining changes in pupil diameter during periods of high and low emotional arousal provides a distinctive insight into the dynamic relationship between physiological responses. Also, differentiating episodes of different levels of arousal reflects assumptions from Yerkes-Dodson Law (Yerkes & Dodson, 1908) and can provide evidence on the assumed interaction of emotional arousal and cognition in this model. Acknowledging an intra-individual variability in the relationship between emotional arousal and pupil diameter, it is posited that the most significant correlations may emerge specifically at the peaks of emotional intensity. Therefore, differentiating episodes characterized by high, medium and low emotional arousal is of interest. Above that, for educational research and practice, also the time lag between the two data streams is crucial.

**RQ2:** Does a systematic and significant time lag exist between the onset of emotional arousal and corresponding changes in pupil diameter measurements?

Third, assessing the time lag between emotional arousal based on facial expression analysis and physiological measures like pupil diameter is crucial, especially for future adaptive learning systems. In real-world scenarios, emotional responses might not be immediate, and there could be time lags between emotional arousal and physiological reactions. In general, emotional responses typically occur before physiological responses. When a person experiences an emotion, such as fear, joy, anger, or surprise, the emotional response is reflected in activation in associated brain areas and triggers a series of physiological changes in the body, such as increased heart rate or changes in skin conductance (sweating), or as outlined pupil diameter (Šimić et al., 2021). By assessing the mean time-lag between emotional arousal and pupil diameter, adaptive learning systems can provide more timely and contextually appropriate feedback or intervention. This adds up to the call that indicators need to be aligned in a way that all indicators are well reflected (Mu et al., 2022). As described above, when merging data in educational practice there is a risk for an analytic bottle neck which can result in latencies in delivering cues and inferences (Azevedo & Gasavi, 2019). For instance, if a learner shows signs of frustration or disengagement based on facial expressions, physiological measures can validate and reinforce these emotional indicators, prompting the system to offer support or encourage the learner appropriately. Therefore, it is important to know about initial reactions and time lags between different psycho-physiological responses measured by different data streams.

## 3. Methodology

3.1 Participants

Requirements and criteria for participants' inclusion in the study were normal or corrected to normal vision and sufficient German language and reading skills to process the emotional dilemma task. Beyond that no criteria for exclusion were defined. In summary, $n = 36$ participants took part in the study. Due to technical issues (e.g. hardware failure when recording videos for emotion recognition, loss of connection between devices between devices due to unstable internet connection) $n = 6$ participants were excluded from the analysis. Two additional participants were excluded during the pre-processing of the data streams. This exclusion was due to extreme outliers concerning the time taken for task execution and substantial gaps in the data, which could not be rectified through standard imputation techniques or outlier management strategies. The final sample was $n = 28$ participants. All participants gave their consent for the multimodal data collection after being informed about the conditions of participation.

Among the participants, $n = 13$ (46.43%) indicated to be male and $n = 15$ (53.57%) to be female. $N = 8$ (cumulative percentage: 28.57%) reported a secondary school or university entry degree to be their highest educational degree certificate, $n = 14$ (50%) a Bachelor's degree, and $n = 6$ (21.43%) a Master's degree. Because the study uses dilemma from work contexts as an emotionally engaging task, participants were also asked for their work experience using a multiple-choice item: $n = 14$ (50%) participants completed vocational training, $n = 6$ (21.34%) completed an internship for longer than six months, $n = 9$ (32.14%) worked part-time during their studies and n = 17 (60.76%) were fully employed for longer than one year. $N = 2$ (7.14%) participants indicated no work experience. Based on the educational degree and the work experience, the sample can be considered as suitable for the emotional dilemma task described in the following.

3.2 Emotional Dilemma Task

In this study, we used three everyday moral dilemma tasks containing a moral conflict as described in Christensen and Gomila (2012) as an emotionally engaging task. The topic and content of the moral dilemmas were chosen with regard to findings that work events like goal or role conflicts are cognitively challenging and can trigger emotions (e.g. Fisher, 2000). The dilemma task was designed as text-based (Dilemma 1: 336 words, dilemma 2: 317 words, dilemma 3: 248 words) and structured as outlined in the following.

Participants of the study were asked to put themselves in the situation described in the dilemmas. The emotional dilemma tasks were structured in an outline of the situation (paragraph 1), a description of a conflict (paragraph 2), and the consequences of different courses of action to solve the dilemma (paragraphs 3 and 4). After gathering the information by reading the text, the participant was asked to decide between two courses of action and justify their decision by three to five arguments. In terms of content, the dilemmas presented either role-

, goal or between-person-conflicts which show the potential to trigger emotions (Fisher, 2000) and can obtain moral conflicts (Christensen & Gomila, 2012). The first dilemma presented a moral conflict in an academic group work setting and the second and third dilemma presented work conflicts in an occupational setting.

The dilemma was validated by asking the participants about their perception of the realism of the dilemma, the occurrence of an inner conflict, the seriousness of the conflict for the protagonist, and the difficulty of decision-making (operationalized as the opportunity to make a decision after a short moment of thought) using a 5-point rating scale (1 = Not at all; 5 = Highly applicable). The validation questions were adapted from the work of Heinrichs & Schadt (2019). The difficulty of the emotional dilemma task is also reflected when taking into account the decision of the participants (Dilemma 1: 35% option 1 and 65% option 2; Dilemma 2: 56% option 1 and 44% option 2; Dilemma 3: 41% option 1 and 59% option 2). Based on these findings, the emotional dilemma tasks are relatively comparable and assumed to be challenging and realistic enough to study the correlation between pupil size and emotional arousal.

Table 1. *Descriptive Data of Validation Questions for Dilemma 1-3*

|  | Dilemma 1 | | Dilemma 2 | | Dilemma 3 | |
| --- | --- | --- | --- | --- | --- | --- |
|  | *M* | *SD* | *M* | *SD* | *M* | *SD* |
| Realism of dilemma | 3.85 | 1.35 | 3.18 | 1.42 | 3.56 | .79 |
| Seriousness of the conflict | 3.24 | 1.18 | 3.65 | 1.28 | 3.35 | 1.20 |
| Manifestation of the inner conflict | 3.68 | 1.27 | 4.18 | 1.03 | 3.41 | 1.21 |
| Difficulty of decision making[a] | 3.15 | 1.26 | 3.94 | 1.20 | 3.26 | 1.14 |

Note. [a]Reverse coded item

3.3 Apparatus

In our lab, we used state-of-the-art equipment and software to collect eye-tracking data and analyze emotional recognition. The primary eye-tracking tool was Tobii Pro Spectrum (Tobii Pro AB, 2014). This device allowed us to track both eyes simultaneously, ensuring comprehensive data collection. The eye tracker was set up as a desktop mount to provide a stable and controlled environment for the participants. To increase the quality of the gaze data, we ask participants to maintain minimal head movement and a constant viewing distance (~55cm). For the calibration process, we implemented a standard multi-point procedure to ensure the accuracy of eye tracking data. This step was critical in matching the eye tracker to the individual characteristics of the participants, thereby improving the reliability of the data. For this study, we were mainly interested in exporting pupil dilation data for each participant. Eye tracking data was sampled with 120 Hz. In addition to the eye tracker, we used Noldus FaceReader (Noldus, 2021) for emotion recognition, which allowed us to correlate eye movement data with emotional responses. The FaceReader 9 used video material that was sampled with 5 Hz and simultaneously with the eye tracking data via a Logitech HD 1080p webcam.

3.4 Data Analysis
*3.4.1 Data Pre-processing*

Data pre-processing is an essential phase when aiming to synchronize pupil diameter with emotional arousal (Fink et al., 2023; Kret & Sjak-Shie, 2018). Data processing and analysis were conducted using Python, leveraging its robust ecosystem of libraries to handle a range of tasks; Pandas (McKinney, 2010), SciPy (Virtanen et al., 2020) and NumPy (Harris et al., 2020). We preprocessed all dilemma scenarios and the corresponding data individually. The preprocessing pipeline is shown in more detail:

1. To address the disruptions caused by blinks in our pupil diameter streams, we implemented a systematic threshold-based correction procedure. As described by Hollander and Huette (2022) there is no consensus on the best way to determine the most appropriate lower blink threshold parameter. We followed the protocol from Huette (2016); blinks were primarily identified based on a minimum blink threshold, which refers to the amount of time that consecutive samples must be missing for an event to be labeled as a blink (here 100ms).

2. Once blinks were detected, our next task was to interpolate and correct the affected data points. For this, we employed cubic spline interpolation, a method well-suited for ensuring a smooth and natural transition between data points (Dyer & Dyer, 2001). To ensure the reliability of our interpolation, we selected three data points immediately preceding the blink (pre-blink) and three data points following the blink (post-blink). These points served as "anchors", enabling the interpolation to be rooted in actual recorded values and ensuring that the interpolated section seamlessly integrated with the genuine data.

3. Downsampling is a technique used to reduce a signal's sampling rate. One prevalent method involves averaging *N* consecutive samples (Fink et al., 2023; Kret & Sjak-Shie, 2018). This strategy is especially advantageous for signals with noise, as it offers a smoother representation by counteracting random fluctuations. The value of *N* directly influences the resulting sampling rate. In our case, given the disparity in sampling rates between pupil data (120Hz) and emotion data (5Hz), we adjusted the gaze data to match the emotion data's 5Hz rate. This was achieved by averaging every 24 consecutive samples. The process begins by segmenting the signal into non-overlapping chunks, each containing 24 samples. Subsequently, the mean of each segment is computed, yielding the downsampled value for that portion of the data. As a result of the downsampling procedure, we have both pupil diameter and emotional arousal in the data set with a sampling rate of 5Hz.

4. Ensuring accurate temporal alignment of two data streams is vital when studying the relationship between pupil diameter and emotional arousal (see Figure 1). The first step in this process involved an examination of the initial timestamps from both data streams. As they were synchronized in the Observer XT software (Zimmerman et al., 2009), they commenced at the identical millisecond. We then adjusted the timestamps so that each participant's data starts at 0 milliseconds. Upon examination of the data, the synchronization proved successful, with no discernible offset detected.

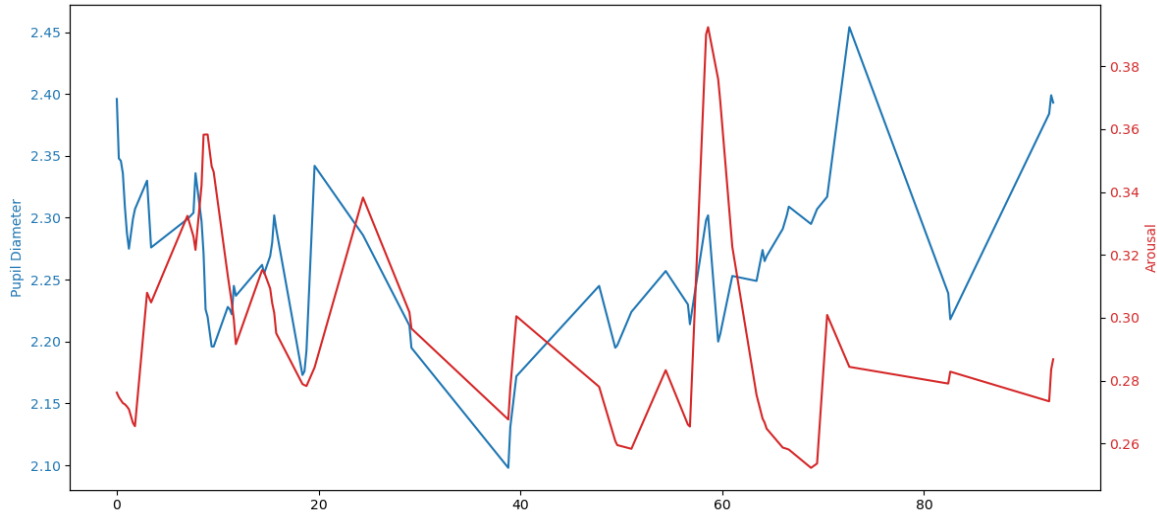

Figure 1. Downsampled and Synchronized Data Streams Over Time (in Sec.): Pupil Diameter Plotted Against Emotional Arousal of a Random Participant

*3.4.2 Data Analysis*

Analyzing the correlation and exploring the time lag between pupil diameter and emotional arousal data streams involves several analysis steps:

1. We commenced our analysis by utilizing the full normalized and interpolated data streams for each participant, upon which we calculated the Spearman correlation co-efficients to evaluate the relationship between emotional arousal and pupil diameter (RQ 1a).

2. Then we explored epochs of high, medium, and low arousal and the corresponding Spearman correlation coefficients between these different epochs and pupil diameter. First, we defined a high in arousal as any epochs of local maximum that were in the upper third quantile (> 66th percentile), epochs of medium arousal in the middle third quantile (between 33rd and 66th percentiles), and epochs of low arousal in the lower third quantile (< 33rd percentile). Within these epochs, we again calculated the Spearman rank correlation between arousal and pupil diameter, providing a non-parametric measure of association that does not assume a linear relationship (RQ 1b).

3. In order to identify significant time lags between both data streams (RQ 2), we applied the so-called Granger causality (Granger, 1969). Granger causality is a statistical concept of causality that is based on prediction. Named after the Nobel laureate Clive Granger, it operates on the principle that if a signal *X* "Granger-causes" a signal *Y*, then past values of *X* should contain information that helps predict *Y* beyond the information contained in past values of *Y* alone. Granger causality tests are particularly insightful when examining the specific lags at which one time-series may forecast another. In essence, the test evaluates whether past values of a time-series *X* at various lags contribute unique information that can predict the current value of another time series *Y*. For instance, if we find that *X* Granger-causes *Y* at lag 2, it suggests that information from *X* two time units ago provides significant predictive power for the current value of *Y*. This lagged relationship can uncover dynamics that are not immediately apparent, revealing how effects propagate over time in a system. The identification of these specific lags is crucial because it pinpoints the temporal distance over which the predictive relationship holds, which can be vital for understanding the underlying processes or for developing forecasting models. It is important to clarify that Granger causality should not be interpreted in the same sense as causality in a physical or deterministic sense; it is better understood as a "predictive causality".

## 4. Results

On average, participants were engaged in the dilemmas $M_{d1}$ = 55.04 sec (*SD* = 36,24), $M_{d2}$ = 39.35 sec (*SD* = 17.30) and $M_{d3}$ = 42.09 sec (*SD* = 31.14). The descriptive data of pupil diameter and emotional arousal are presented in Table 2.

*Table 2. Descriptive Data of Pupil Diameter and Emotional Arousal separated by Dilemma (D1, D2, D3)*

|  | **Mean** | **SD** | **Min.** | **Md.** | **Q3** | **Max.** |
|---|---|---|---|---|---|---|
| Pupil Diameter |  |  |  |  |  |  |
| D1 | 2.54 | .25 | 1.73 | 2.53 | 2.70 | 3.48 |
| D2 | 2.57 | .24 | 1.85 | 2.57 | 2.69 | 4.21 |
| D3 | 2.59 | .24 | 2.05 | 2.55 | 2.70 | 3.39 |
| Emotional Arousal |  |  |  |  |  |  |
| D1 | .28 | .04 | .01 | .27 | .29 | .77 |
| D2 | .29 | .06 | .02 | .26 | .29 | .74 |
| D3 | .29 | .05 | .05 | .28 | .30 | .68 |

*Note*. Q3 indicates mean values for the upper third quartile (>66[th] percentile); values for emotional arousal ranged from 0 to 1

## 4.1 Correlation between Pupil Diameter and Emotional Arousal (RQ1a)

To assess the distribution characteristics of emotional arousal and pupil diameter across participants, we employed Shapiro-Wilk tests. Our findings revealed significant deviations from a normal distribution for both variables, indicated by p-values below the .05 threshold. As a consequence, adopting non-parametric methods for subsequent analyses is imperative, aligning with recommendations by Fink et al. (2023). Consequently, we utilized Spearman rank correlation coefficients to investigate the relationship between emotional arousal and pupil diameter. Our analysis yielded the following insights, categorized by Dilemmas 1-3.

For the first dilemma (D1), analysis revealed a mean Spearman correlation coefficient of -.13 across the dataset, indicating a generally weak inverse relationship between the two data streams, which was statistically significant at a p-value of less than .001. This correlation coefficient varied significantly among the participants, with the strength of negative correlations reaching as low as -.65 and positive correlations peaking at .38. This wide range underscores the significant individual variability in the relationship between emotional arousal and changes in pupil diameter, suggesting that the connection between these data streams is not only variable but also highly individualized.

For the second dilemma (D2), the Spearman correlation coefficient was found to be -.09, with the data showing a range from -.42 to .22. Despite being weaker than the correlation observed in the first dilemma (D1), these results still indicate a significant negative correlation, supported by a p-value of less than .001, and also highlight a high level of variability among the participants.

Regarding the third dilemma (D3), the analysis yielded a mean Spearman correlation coefficient of -.11, with correlations spanning from -.53 to .25. This outcome, akin to D1, points to a moderate negative association between the variables, which is statistically significant ($p < .001$), underscoring the consistent presence of a significant yet variable inverse relationship across different dilemmas.

Using Spearman rank correlation coefficients, we observed consistent and statistically significant overall negative associations between emotional arousal and pupil diameter across all three dilemmas. Notably, the range of correlation coefficients was quite large, suggesting variability in the strength of the relationship across and within individuals. Therefore, further investigation is needed to elucidate the underlying factors contributing to this variability. One of these further investigations is to examine the relationship between pupil diameter and emotional arousal in different epochs of high, medium, and low emotional arousal.

## 4.2 Correlation of Pupil Diameter and Emotional Arousal in Epochs of High, Medium and Low Emotional Arousal (RQ1b)

We then implemented a quantile-based thresholding approach to identify highs in the emotional arousal data stream. Using a quantile-based thresholding approach to classify arousal states into high, medium, and low epochs reveals a balanced distribution among participants across the three dilemmas, with only marginal variations observed between them. This approach ensures that the categorization of arousal levels is directly tied to the data's distribution, making it inherently balanced. Such a method divides the arousal scores into equal-sized groups based on their distribution, which naturally leads to a balanced number of observations across the defined categories (high, medium, low). The balanced distribution suggests that the dilemmas were effective in eliciting a wide range of emotional responses, with no single arousal epoch dominating across the scenarios.

The Spearman rank correlation coefficient between high emotional epochs (defined as epochs above the 66th percentile) and pupil size is -.13 ($p<.001$) in D1, -.18 in D2 ($p<.001$), and -.16 in D3 ($p<.001$). These results show a significant, moderate negative correlation between heightened arousal highs and pupil diameter in all three dilemmas, suggesting that heightened arousal levels are moderately associated with smaller pupil sizes.

Our investigation then turned to epochs characterized by medium emotional arousal. The Spearman rank correlation coefficients for these epochs (defined as falling between the 33rd and 66th percentiles) and corresponding pupil sizes are approximately D1 = .08 ($p<.11$), D2 = .04 ($p<.09$), and D3 = .09 ($p<.14$). These results suggest a slightly positive but not statistically significant correlation.

We then conducted a parallel investigation focusing on epochs of low emotional arousal. The Spearman rank correlation coefficients for these epochs (defined as falling below the 33rd percentile) and pupil size are approximately D1 = .07 ($p<.12$), D2 = .04 ($p<.13$), and D3 = .10 ($p<.09$). Similar to the moderate arousal epochs, these results indicate a slightly positive but not statistically significant correlation.

These results suggest variability in the relationship between arousal and pupil diameter across different arousal levels. Specifically, in all three dilemmas a transition from positive correlations in lower arousal states to highly significant negative correlations in higher arousal states is found.

## 4.3 Time Lags Between the Onset of Emotional Arousal and Corresponding Changes in Pupil Diameter Measurements (RQ 2)

Granger causality is a statistical hypothesis test to determine whether one time series can predict another time series. To calculate the mean time lag between the two time-series streams for each participant, we need to determine the specific time lags at which the Granger causality is significant for each participant. A significant Granger causality result indicates a predictive relationship between the two time-series (see *p*-values in Table 3-5). If the emotional arousal stream Granger-causes the pupil diameter stream, it suggests that past values of emotional arousal contain information that is useful in predicting future values of pupil diameter.

*Table 3. Granger Causality for Emotional Arousal Effects Pupil Diameter at Different Time Lags in Dilemma 1*

| ID | Lag 1 | Lag 2 | Lag 3 | Lag 4 | Lag 5 | Lag 6 | Lag 7 | Lag 8 | Lag 9 | Lag 10 |
|---|---|---|---|---|---|---|---|---|---|---|
| 1 | **.003** | — | — | — | — | — | — | — | — | — |
| 2 | .22 | .62 | .79 | .71 | .81 | .20 | .07 | .10 | .17 | .16 |
| 3 | .44 | .99 | .88 | .81 | .70 | .69 | .83 | .74 | .81 | .87 |
| 4 | **.05** | — | — | — | — | — | — | — | — | — |
| 5 | .46 | .24 | .19 | .15 | **0.05** | — | — | — | — | — |
| 6 | **.05** | — | — | — | — | — | — | — | — | — |
| 7 | 0.06 | **<.001** | — | — | — | — | — | — | — | — |
| 8 | .11 | .05 | **.05** | — | — | — | — | — | — | — |
| 9 | .28 | .71 | .75 | .40 | .59 | .56 | .75 | .24 | .26 | .28 |
| 10 | .12 | .51 | .83 | .88 | .61 | .73 | .91 | .94 | .93 | .91 |
| 11 | .63 | **.01** | — | — | — | — | — | — | — | — |
| 12 | .26 | .34 | .41 | .06 | .06 | **.05** | — | — | — | — |
| 13 | **<.001** | — | — | — | — | — | — | — | — | — |
| 14 | .44 | .74 | .88 | .95 | .97 | .98 | .99 | .98 | .99 | .99 |
| 15 | .33 | .49 | .62 | .69 | .81 | .87 | .92 | .92 | .93 | .96 |
| 16 | **.002** | — | — | — | — | — | — | — | — | — |
| 17 | .84 | .98 | .90 | .79 | .60 | **.04** | — | — | — | — |
| 18 | .13 | .39 | .44 | .41 | .58 | .40 | .48 | .49 | .37 | .43 |
| 19 | **.05** | — | — | — | — | — | — | — | .21 | .29 |
| 20 | .19 | 0.75 | .89 | .83 | .89 | .93 | .87 | .91 | .33 | **.02** |
| 21 | **.05** | — | — | — | — | — | — | — | — | — |
| 22 | **.05** | — | — | — | — | — | — | — | — | — |
| 23 | **.05** | — | — | — | — | — | — | — | — | — |
| 24 | .51 | .57 | .75 | .86 | .95 | .91 | .94 | .97 | .99 | .97 |
| 25 | .09 | **<.001** | — | — | — | — | — | — | — | — |
| 26 | .42 | .36 | .06 | .07 | .11 | .21 | .40 | .45 | .29 | .34 |
| 27 | .96 | .07 | .14 | .58 | .55 | .30 | .35 | .29 | .31 | .34 |
| 28 | .07 | .93 | .84 | .47 | .15 | .22 | .24 | .22 | .15 | .20 |

Note. *p*-values of Granger-causality tests for every participant. Cells are replaced with "—" after a lower time lag was significant for the corresponding participant.

The mean time lag for the first dilemma, across all participants who showed significant Granger causality, is approximately 2.81 (or 2.81 ms). This value represents the average lag at which the relationship between arousal and pupil diameter becomes significant, indicating the typical delay in the effect across participants.

*Table 4. Granger Causality for Emotional Arousal Effects Pupil Diameter at Different Time Lags in Dilemma 2*

| ID | Lag 1 | Lag 2 | Lag 3 | Lag 4 | Lag 5 | Lag 6 | Lag 7 | Lag 8 | Lag 9 | Lag 10 |
|---|---|---|---|---|---|---|---|---|---|---|
| 1 | .03 | — | — | — | — | — | — | — | — | — |
| 2 | .76 | .96 | .48 | .59 | .42 | .23 | .33 | .44 | .53 | .62 |
| 3 | .26 | .50 | .58 | .75 | .94 | .96 | .96 | .82 | .67 | .67 |
| 4 | .46 | .37 | .58 | .50 | .89 | .95 | .87 | **<.001** | — | — |
| 5 | **0.05** | — | — | — | — | — | — | — | — | — |
| 6 | .28 | .36 | .62 | .76 | .80 | .73 | .89 | .90 | .67 | .67 |
| 7 | **<.001** | — | — | — | — | — | — | — | — | — |
| 8 | .98 | .20 | .73 | .65 | .65 | .70 | **.03** | — | — | — |
| 9 | .22 | **.01** | — | — | — | — | — | — | — | — |
| 10 | .35 | .65 | .57 | .86 | .93 | .92 | .93 | .85 | .60 | .69 |
| 11 | **.03** | — | — | — | — | — | — | — | — | — |
| 12 | **.01** | — | — | — | — | — | — | — | — | — |
| 13 | .56 | .72 | .77 | .46 | .54 | .66 | .62 | .63 | .76 | .54 |
| 14 | .26 | .30 | .42 | .28 | .33 | .29 | .14 | **.03** | — | — |
| 15 | **<.001** | — | — | — | — | — | — | — | — | — |
| 16 | **.01** | — | — | — | — | — | — | — | — | — |
| 17 | .60 | **<.001** | — | — | — | — | — | — | — | — |
| 18 | .62 | .69 | .75 | .49 | .22 | .06 | .12 | .13 | .14 | .16 |
| 19 | .55 | .15 | **.03** | — | — | — | — | — | — | — |
| 20 | **.02** | — | — | — | — | — | — | — | — | — |
| 21 | .15 | .31 | .34 | .26 | .18 | .52 | **.03** | — | — | — |
| 22 | .88 | .61 | .71 | .47 | .21 | .26 | .11 | .08 | .12 | .17 |
| 23 | **<.001** | — | — | — | — | — | — | — | — | — |
| 24 | **.02** | — | — | — | — | — | — | — | — | — |
| 25 | .76 | .96 | .48 | .59 | .42 | .23 | .33 | .44 | .53 | .62 |
| 26 | .26 | .50 | .58 | .75 | .94 | .96 | .96 | .82 | .67 | .67 |
| 27 | .46 | .37 | .58 | .50 | .89 | .95 | .87 | **<.001** | — | — |
| 28 | .08 | .71 | .88 | .93 | .98 | .98 | .99 | .89 | .88 | .78 |

Note. *p*-values of Granger-causality tests for every participant. Cells are replaced with "—" after a lower time lag was significant for the corresponding participant.

The mean time lag for the first dilemma, across all participants who showed significant Granger causality, is approximately 3.44 (or 3.44 ms).

*Table 5. Granger Causality for Emotional Arousal Effects Pupil Diameter at Different Time Lags in Dilemma 3*

| ID | Lag 1 | Lag 2 | Lag 3 | Lag 4 | Lag 5 | Lag 6 | Lag 7 | Lag 8 | Lag 9 | Lag 10 |
|---|---|---|---|---|---|---|---|---|---|---|
| 1 | .60 | .11 | .20 | .12 | **.02** | — | — | — | — | — |
| 2 | .77 | .68 | .85 | .58 | .36 | .24 | .29 | .30 | .33 | .41 |
| 3 | **<.001** | — | — | — | — | — | — | — | — | — |
| 4 | .62 | .79 | .91 | .90 | .49 | .15 | .19 | .21 | .09 | .09 |
| 5 | .28 | .21 | .24 | .34 | .47 | .48 | .28 | .37 | .32 | .33 |
| 6 | .46 | .54 | .78 | .90 | .88 | .63 | .71 | .77 | .85 | .84 |
| 7 | .12 | .91 | .79 | .64 | .36 | .43 | .09 | .10 | .16 | .21 |
| 8 | .30 | .63 | .55 | .17 | .07 | .09 | .12 | .16 | .22 | .20 |
| 9 | .72 | .33 | .47 | .58 | .27 | .13 | .19 | .26 | .32 | .31 |
| 10 | .80 | .89 | .19 | .08 | .12 | .19 | .22 | .28 | .16 | .14 |
| 11 | .98 | .61 | .71 | .90 | .89 | .75 | .85 | .28 | .06 | .05 |
| 12 | .61 | .67 | .76 | **.05** | — | — | — | — | — | — |
| 13 | .75 | .44 | .30 | .51 | .56 | .29 | .37 | .45 | .55 | .64 |
| 14 | .47 | **<.001** | — | — | — | — | — | — | — | — |
| 15 | .07 | .52 | .65 | .80 | .88 | .60 | .67 | .40 | .37 | .45 |
| 16 | **<.001** | — | — | — | — | — | — | — | — | — |
| 17 | **.05** | — | — | — | — | — | — | — | — | — |
| 18 | **<.001** | — | — | — | — | — | — | — | — | — |
| 19 | .89 | .20 | .38 | .51 | .60 | .60 | .44 | .51 | .70 | .76 |
| 20 | **.02** | — | — | — | — | — | — | — | — | — |

| | | | | | | | | | |
|---|---|---|---|---|---|---|---|---|---|
| 21 | **.03** | — | — | — | — | — | — | — | — |
| 22 | .30 | .85 | .60 | .30 | .26 | .35 | .50 | .59 | .63 | .71 |
| 23 | .86 | .65 | .66 | .80 | .90 | .83 | .89 | .89 | .78 | .20 |
| 24 | .18 | .41 | .64 | .79 | .73 | .81 | .85 | .91 | .93 | .96 |
| 25 | .14 | .65 | .83 | .94 | .93 | .99 | .96 | .96 | .76 | .85 |
| 26 | .39 | **.04** | — | — | — | — | — | — | — | — |
| 27 | .33 | .97 | 1,00 | .90 | .84 | .91 | .81 | .66 | .21 | .21 |
| 28 | .60 | .11 | .20 | .12 | **.02** | — | — | — | — | — |

Note. *p*-values of Granger-causality tests for every participant. Cells are replaced with "—" after a lower time lag was significant for the corresponding participant.

The mean time lag for the first dilemma, across all participants who showed significant Granger causality, is approximately 2.18 (or 2.18 ms).

## 5. Discussion

This study investigates the synchronization and correlation of pupil diameter, measured using a stationary eye tracker, and emotional arousal, measured through facial expression recognition, during emotional dilemma tasks. We utilized three everyday moral dilemma scenarios characterized by moral conflicts, chosen based on previous literature indicating their capacity to elicit emotions and cognitive demands, as observed in similar studies (e.g., Fisher, 2000). Our aim was to unveil potential correlations and explore time lags between these psychophysiological data streams. Analysis revealed moderate negative associations between emotional arousal and pupil diameter across all three emotional dilemmas. Additionally, heightened emotional arousal was notably correlated with smaller pupil size, while moderate and low arousal epochs showed slightly positive but non-significant correlations with pupil diameter. These findings underline the nuanced relationship between arousal and pupil diameter across varying arousal levels. We proceeded to investigate time lags, assessing if one data stream reliably predicts the other while considering (i.e., simulate) different time lags, using Granger causality analysis. Our findings revealed an average lag of approximately 2.81 ms, at which the relationship between arousal and pupil diameter becomes significant. This indicates a typical delay in the effect observed across participants.

5.1 Pupil Diameter and Emotional Arousal Correlate During Epochs of High Emotional Arousal

Our research delves into the realm of multimodal approaches that integrate cognitive and emotional indicators, filling a gap identified by Noorozi et al. (2020). Inspired by the work of Mu et al. (2022), we explore the benefits of multimodal data validation by analyzing correlations between data streams, challenging their concurrent and convergent validity. We found a significant negative correlation between pupil dilation and emotional arousal systematically across all three moral dilemmas. This finding seems to contradict the reported phenomenon on an aggregated level that increased pupil size (as an indicator for cognitive load) goes along with states of higher emotional arousal (Bradley et al., 2008; Partala & Surraka, 2003; Babiker et al., 2013). Nevertheless, it need to emphazised, that these discrepancies can also derive from different methodological differences as a variety of previous studies choose a approaches such as measuring one indicator during experimental manipulations and emotion inducations instead of multimodal measurements. In our study, we focused on analyzing the correlation between two data streams *over time* as indicators for emotional arousal and cognitive load, which is an essential part of multimodal data validation (Mu et al., 2020). However, while the majority of the above listed research suggests a positive correlation between pupil diameter and emotional arousal, indicating that pupil diameter tends to increase with increasing emotional arousal, there are instances where a negative correlation or no correlation has been observed (Hess & Polt, 1960; Bebko et al., 2011). One reason for this discrepancy could be attributed to the everyday moral dilemma embedded within our experimental design. In our study, we assessed both pupil diameter and emotional arousal while participants were actively engaged in moral dilemmas, requiring them to employ strategies to regulate their emotions. This complex interplay between moral judgement and emotional regulation likely influences the observed patterns in pupil responses. For example, Bebko et al. (2011) observed declining pupil diameter during the process of emotion regulation. They observed that when individuals engage in strategies to regulate their emotions, such as suppressing negative emotions, the pupil size tends to decrease. On the other hand, other studies have found that when individuals use reappraisal to regulate their emotions (such as reframing a negative situation in a more positive light), both increasing and decreasing negative emotions can lead to pupil dilation. This dilation is likely a result of increased cognitive effort required to regulate emotional responses through reappraisal. So, why the apparent discrepancy? One possible explanation is that different emotion regulation strategies may have distinct effects on pupil size. For instance, reappraisal (in comparison to suppression) involves cognitive restructuring and may require more cognitive effort, leading to larger pupil sizes (Bebko et al., 2011). Moreover, we posit that the variance (between participants) in the correlation between emotional arousal and pupil diameter could be also attributed to the utilization of different emotion regulation strategies. Our findings indicate that since we also observed positive correlations between arousal and pupil diameter for some participants. This may suggest that distinct regulation strategies, such as reappraisal versus suppression, may influence the direction and strength of the correlation. For instance, while suppression of emotions may lead to decreased pupil size due to the inhibition of emotional expressions, reappraisal may entail increased cognitive effort, resulting in larger pupil sizes. Therefore, the participants' emotion regulation strategy could contribute to the observed variability in the relationship between arousal and pupil diameter. This line of reasoning is also supported by research by Kinner and colleagues (2017), which suggests that pupil diameter is modulated by emotional arousal, but that it is initially related to the amount of mental effort required to regulate automatic emotional responses.

Also, early research conducted by Stanners and colleagues (1979), as well as more recent findings by Chen and Epps (2013), have demonstrated that the pupil dilates in response to emotional arousal primarily when cognitive task demands are minimized. These studies suggest that cognitive processes may exert greater control over the pupillary response in tasks that involve both cognitive and emotional components. These previous studies further support our results, as we have shown the absence of any correlation between data streams during epochs characterized by low or moderate emotional arousal states. This, furthermore, also aligns closely with the principles of the Yerkes-Dodson Law, which posits that excessively high levels of emotional arousal might impede cognitive processes crucial for learning (Sherwood, 1965; Yerkes & Dodson, 1908). Considering that pupil diameter is an indicator for cognitive load (Mallick et al., 2016; Soussou et al., 2012; Rodemer et al., 2023), high arousal might lead to reduced pupil diameters because it affects working memory and reduces capability for spending effort in cognitively demanding tasks. It is also in line with findings showing that stress can hinder learning and stressed individuals show attenuated pupillary responses (de Berker et al., 2016). Also, research on moral judgement indicates that strong emotional triggers lead to more automatic-emotional processes instead of deliberate-controlled processes typically associated with cognitive load (Greene et al., 2001, 2009). In summary, it can be concluded that a closer analysis of the correlation of the psycho-physiological data streams in intervals with different levels of arousal is a useful addition to the current state of research.

In addition to the influence of emotion regulation strategies on pupil size and cognitive information processing, it is essential to consider the impact of emotional valence on this relationship (Kinner et al., 2017; Partala & Surraka, 2003; Babiker et al., 2013; Alsheri & Alghowinem, 2013). Emotional valence, or the positivity or negativity of an emotion, can significantly influence pupillary responses. For instance, Hess and Polt's (1960) early discovery of a bi-directional effect remains influential in the study of pupil responses to emotion. Both Hess and Polt (1960) and Kinner et al. (2017) demonstrated that emotional valence can modulate the relationship between pupil diameter and emotional arousal, indicating that the specific emotional content plays a crucial role in shaping pupillary responses. It suggests that the relationship between pupil diameter and emotional arousal may be complex and context dependent, with different emotional states (valence & arousal) eliciting different patterns of pupil responses.

5.2 Emotional Arousal Precedes and Triggers Changes in Pupil Diameter

The methodological framework of our study marks a notable advance by integrating complex Granger causality analysis with psycho-physiological data streams. This innovative approach not only represents a technical advance in psycho-physiological data analysis (Shojaie & Fox, 2022), but also enriches our understanding of the temporal relationships between emotional arousal and changes in pupil diameter. The Granger causality test, a key component of our methodology, is unique in its ability to statistically determine how one time series may predict future changes in another. Thereto, this feature facilitates the identification of directional relationships between time series, providing a solid foundation for revealing the influence of one variable on another (Barnett et al., 2009). Through this analytical lens, we are able to delve deeper into the dynamics at play, providing a clearer picture of the pathways underlying the physiological manifestations of emotional processes and, therefore, extending the literature systematically (Oliva & Anikin, 2018). We can confirm research utilizing a diverse array of psycho-physiological sensor streams has consistently shown that within the broad spectrum of physiological reactions, emotional arousal often precedes alterations in pupil diameter (Bradley et al., 2008, 2017; Oliva & Anikin, 2018). This body of research highlights that emotional arousal stimulates the sympathetic nervous system, leading to a range of physiological effects, including increased heart rate, elevated perspiration levels, and variations in pupil size, with our study concentrating specifically on the latter. Notably, pupil dilation is directly linked to increased sympathetic nervous system activity, whether in response to external stimuli or arising spontaneously. Consequently, our results lend further support to the hypothesis that emotional arousal initiates changes in pupil diameter, aligning with the findings of prior research such as Bradley et al. (2008). Uniquely, to our knowledge, our study is the first to apply an emotion recognition system that captures facial expressions to produce a continuous stream of emotional arousal data for the purpose of correlating this stream with physiological data stream such as pupil diameter. Previous methodologies required participants to manually indicate their arousal or valence levels using button presses (Child et al., 2020; Kinner et at., 2017; Oliva & Anikin, 2018). Our approach, offering millisecond temporal resolution of the emotional arousal stream, not only allows for a precise correlation between emotional arousal and pupil diameter changes but also confirms that the emotional arousal precedes these changes. Moreover, our analysis did not uncover any significant Granger causality in the reverse direction, suggesting that pupil diameter changes follow rather than lead emotional arousal. This absence of reverse causality strongly supports the notion that emotional arousal is a primary driver of pupil diameter changes, rather than the converse.

It is important to note that our results indicate that Granger causality analysis did not yield statistically significant results for each participant when considering time lags of up to 10 milliseconds. This observation suggests a complex interplay between emotional arousal and pupil diameter changes that may not be captured consistently across individuals within such a short time window. This lack of consistency across participants underscores the inherent variability in psychophysiological responses to emotional stimuli, reflecting individual differences in the rate and manner of processing such stimuli. It suggests the potential need for a more detailed analysis, or possibly the inclusion of additional data points or variables, that could help to better understand these dynamics. Furthermore, this finding encourages further investigation into the temporal resolution and sensitivity of our methods when exploring the intricate causal relationships within psychophysiological data streams.

For educational practice and research, it is important to consider possible analytic bottlenecks which can occur when a high variety of multimodal data streams are merged. As described by Azevedo & Gasevic, latencies in delivering inferences and cues caused by this analytic bottleneck might negatively affect the learning process. Therefore, it is crucial to to be aware of congruent vadility of measures to select the right data streams and combine them in a beneficial way (Mu et al., 2022) and use the knowledge of time lags to identify and use "initial reactions" for cues in adaptive and remote learning environments.

5.3 Limitations & Future Research

While this investigation has shed light on the complex relationship between emotional arousal and changes in pupil diameter, it is important to acknowledge several limitations that may inform future research directions. First, our study did not consider the potential influence of individual differences in emotion regulation strategies, which are known to significantly influence physiological responses to emotional stimuli. It is critical that future studies assess participants' emotion regulation strategies to disentangle how these strategies may play a role in mediating or moderating physiological outcomes in response to emotional arousal.

In addition, our methodology omitted the assessment of participants' stress levels, which may have missed a crucial element that affects both emotional arousal and its physiological responses. Implementing measures to assess stress levels both before and during the experimental procedure could provide critical insight into how stress modulates emotional arousal and its associated physiological responses.

Furthermore, the experimental design did not adequately account for variability in the intensity of the dilemmas presented to participants, which could have led to discrepancies in task load and, by extension, cognitive workload. Cognitive workload, which encompasses both task load and mental effort, likely influenced participants' emotional and physiological responses, but this variable was not methodologically controlled or quantified in our study. Future research should strive for a standardized approach to dilemma or task presentation to ensure consistent cognitive load across different scenarios. In addition, examining how cognitive workload interacts with emotional arousal may shed light on the mechanisms by which these factors collectively influence physiological responses.

In light of our findings (especially RQ1a & RQ1b), future research should seek to enrich our understanding of the nuanced relationship between emotional arousal and physiological indicators, such as changes in pupil diameter, paying particular attention to the variability observed among individual participants. The discovery that the correlation between emotional arousal and pupil diameter tends to be slightly below zero, albeit with considerable variability across participants, highlights the need to explore individual differences more deeply. This exploration is crucial for disentangling the layers of complexity that characterize the interplay between emotional processes and physiological responses

5.4 Conclusion

This study explores the complex relationship between emotional arousal and pupil diameter changes using moral dilemmas to elicit emotional and cognitive responses. By integrating stationary eye-tracking and facial expression recognition technologies, we aimed to identify correlations and time lags between these psycho-physiological data streams. Our results show a moderate negative correlation between emotional arousal and pupil diameter, suggesting a nuanced interaction that varies with arousal intensity. Specifically, higher levels of emotional arousal are associated with smaller pupil size, whereas lower levels of arousal have a less pronounced, non-significant effect on pupil diameter. Further exploration using Granger causality analysis revealed a typical delay of approximately 2.8 ms in the significant relationship between arousal and pupil changes, suggesting a short latency in the physiological response to emotional stimuli across individuals. These results contribute to our understanding of the dynamic interplay between emotional states and physiological responses, and highlight the complexity of the impact of emotional arousal on pupil diameter in cognitive-emotional contexts. Thereby the results emphasize the relevance of research using multimodal approaches and considering the convergent validity of measures for cognitive workload and emotional arousal to improve adaptive learning environments.